# OGLE-2018-BLG-0799Lb: a $q \sim 2.7 \times 10^{-3}$ Planet with Spitzer Parallax

Weicheng Zang[1,A,C,E⋆], Yossi Shvartzvald[2,A,C,E], Andrzej Udalski[3,B], Jennifer C. Yee[4,A,C,E], Chung-Uk Lee[5,6,C], Takahiro Sumi[7,D], Xiangyu Zhang[1], Hongjing Yang[1], Shude Mao[1,8,E], Sebastiano Calchi Novati[9,A], Andrew Gould[10,11,A,C], Wei Zhu[12,A], Charles A. Beichman[9,A], Geoffery Bryden[13,A], Sean Carey[9,A], B. Scott Gaudi[11,A], Calen B. Henderson[9,A], Przemek Mróz[14,B], Jan Skowron[3,B], Radoslaw Poleski[3,B], Michał K. Szymański[3,B], Igor Soszyński[3,B], Paweł Pietrukowicz[3,B], Szymon Kozłowski[3,B], Krzysztof Ulaczyk[15,B], Krzysztof A. Rybicki[3,B], Patryk Iwanek[3,B], Marcin Wrona[3,B], Michael D. Albrow[16,C], Sun-Ju Chung[5,6,C], Cheongho Han[17,C], Kyu-Ha Hwang[5,C], Youn Kil Jung[5,C], Yoon-Hyun Ryu[5,C], In-Gu Shin[5,C], Sang-Mok Cha[5,18,C], Dong-Jin Kim[5,C], Hyoun-Woo Kim[5,C], Seung-Lee Kim[5,6,C], Dong-Joo Lee[5,C], Yongseok Lee[5,18,C], Byeong-Gon Park[5,6,C], Richard W. Pogge[11,C], Ian A. Bond[19,D], Fumio Abe[20,D], Richard Barry[21,D], David P. Bennett[21,22,D], Aparna Bhattacharya[21,22,D], Martin Donachie[23,D], Hirosane Fujii[7,D], Akihiko Fukui[24,25,D], Yuki Hirao[7,D], Yoshitaka Itow[20,D], Rintaro Kirikawa[7,D], Iona Kondo[7,D], Naoki Koshimoto[26,27,D], Man Cheung Alex Li[23,D], Yutaka Matsubara[20,D], Yasushi Muraki[20,D], Shota Miyazaki[7,D], Clément Ranc[21,D], Nicholas J. Rattenbury[23,D], Yuki Satoh[7,D], Hikaru Shoji[7,D], Daisuke Suzuki[29,D], Yuzuru Tanaka[7,D], Paul J. Tristram[30,D], Tsubasa Yamawaki[7,D], Atsunori Yonehara[31,D], Etienne Bachelet[32,E], Markus P.G. Hundertmark[33,E], R. Figuera Jaimes[34,1,E], Dan Maoz[35,E], Matthew T. Penny[36,E], Rachel A. Street[32,E], Yiannis Tsapras[33,E]

[1]*Department of Astronomy and Tsinghua Centre for Astrophysics, Tsinghua University, Beijing 100084, China*
[2]*Department of Particle Physics and Astrophysics, Weizmann Institute of Science, Rehovot 76100, Israel*
[3]*Astronomical Observatory, University of Warsaw, Al. Ujazdowskie 4, 00-478 Warszawa, Poland*
[4]*Center for Astrophysics | Harvard & Smithsonian, 60 Garden St.,Cambridge, MA 02138, USA*
[5]*Korea Astronomy and Space Science Institute, Daejon 34055, Republic of Korea*
[6]*University of Science and Technology, Korea, (UST), 217 Gajeong-ro Yuseong-gu, Daejeon 34113, Republic of Korea*
[7]*Department of Earth and Space Science, Graduate School of Science, Osaka University, Toyonaka, Osaka 560-0043, Japan*
[8]*National Astronomical Observatories, Chinese Academy of Sciences, Beijing 100101, China*
[9]*IPAC, Mail Code 100-22, Caltech, 1200 E. California Blvd., Pasadena, CA 91125, USA*
[10]*Max-Planck-Institute for Astronomy, Königstuhl 17, 69117 Heidelberg, Germany*
[11]*Department of Astronomy, Ohio State University, 140 W. 18th Ave., Columbus, OH 43210, USA*
[12]*Canadian Institute for Theoretical Astrophysics, University of Toronto, 60 St George Street, Toronto, ON M5S 3H8, Canada*
[13]*Jet Propulsion Laboratory, California Institute of Technology, 4800 Oak Grove Drive, Pasadena, CA 91109, USA*
[14]*Division of Physics, Mathematics, and Astronomy, California Institute of Technology, Pasadena, CA 91125, USA*
[15]*Department of Physics, University of Warwick, Gibbet Hill Road, Coventry, CV4 7AL, UK*
[16]*University of Canterbury, Department of Physics and Astronomy, Private Bag 4800, Christchurch 8020, New Zealand*
[17]*Department of Physics, Chungbuk National University, Cheongju 28644, Republic of Korea*
[18]*School of Space Research, Kyung Hee University, Yongin, Kyeonggi 17104, Republic of Korea*
[19]*Institute of Natural and Mathematical Sciences, Massey University, Auckland 0745, New Zealand*
[20]*Institute for Space-Earth Environmental Research, Nagoya University, Nagoya 464-8601, Japan*
[21]*Code 667, NASA Goddard Space Flight Center, Greenbelt, MD 20771, USA*
[22]*Department of Astronomy, University of Maryland, College Park, MD 20742, USA*
[23]*Department of Physics, University of Auckland, Private Bag 92019, Auckland, New Zealand*
[24]*Department of Earth and Planetary Science, Graduate School of Science, The University of Tokyo, 7-3-1 Hongo, Bunkyo-ku, Tokyo 113-0033, Japan*
[25]*Instituto de Astrofísica de Canarias, Vía Láctea s/n, E-38205 La Laguna, Tenerife, Spain*
[26]*Department of Astronomy, Graduate School of Science, The University of Tokyo, 7-3-1 Hongo, Bunkyo-ku, Tokyo 113-0033, Japan*
[27]*National Astronomical Observatory of Japan, 2-21-1 Osawa, Mitaka, Tokyo 181-8588, Japan*
[28]*School of Chemical and Physical Sciences, Victoria University, Wellington, New Zealand*
[29]*Institute of Space and Astronautical Science, Japan Aerospace Exploration Agency, 3-1-1 Yoshinodai, Chuo, Sagamihara, Kanagawa, 252-5210, Japan*
[30]*University of Canterbury Mt. John Observatory, P.O. Box 56, Lake Tekapo 8770, New Zealand*
[31]*Department of Physics, Faculty of Science, Kyoto Sangyo University, 603-8555 Kyoto, Japan*
[32]*Las Cumbres Observatory Global Telescope Network, 6740 Cortona Drive, suite 102, Goleta, CA 93117, USA*
[33]*Zentrum für Astronomie der Universität Heidelberg, Astronomisches Rechen-Institut, Mönchhofstr. 12-14, 69120 Heidelberg, Germany*
[34]*Facultad de Ingeniería y Tecnología, Universidad San Sebastian, General Lagos 1163, Valdivia 5110693, Chile*
[35]*School of Physics and Astronomy, Tel-Aviv University, Tel-Aviv 6997801, Israel*
[36]*Department of Physics and Astronomy, Louisiana State University, Baton Rouge, LA 70803 USA*
[A]*The Spitzer Team*
[B]*The OGLE Collaboration*
[C]*The KMTNet Collaboration*
[D]*The MOA Collaboration*
[E]*The LCO/Spitzer Follow-up Team*

**ABSTRACT**

We report the discovery and analysis of a planet in the microlensing event OGLE-2018-BLG-0799. The planetary signal was observed by several ground-based telescopes, and the planet-host mass ratio is $q = (2.65 \pm 0.16) \times 10^{-3}$. The ground-based observations yield a constraint on the angular Einstein radius $\theta_E$, and the microlensing parallax vector $\vec{\pi}_E$, is strongly constrained by the *Spitzer* data. However, the 2019 *Spitzer* baseline data reveal systematics in the *Spitzer* photometry, so there is ambiguity in the magnitude of the parallax. In our preferred interpretation, a full Bayesian analysis using a Galactic model indicates that the planetary system is composed of an $M_{\text{planet}} = 0.26^{+0.22}_{-0.11} M_J$ planet orbiting an $M_{\text{host}} = 0.093^{+0.082}_{-0.038} M_\odot$, at a distance of $D_L = 3.71^{+3.24}_{-1.70}$ kpc. An alternate interpretation of the data shifts the localization of the minima along the arc-shaped microlens parallax constraints. This, in turn, yields a more massive host with median mass of $0.13 M_\odot$ at a distance of 6.3 kpc. This analysis demonstrates the robustness of the osculating circles formalism, but shows that further investigation is needed to assess how systematics affect the specific localization of the microlens parallax vector and, consequently, the inferred physical parameters.

## 1 INTRODUCTION

Very low-mass (VLM; $M \leq 0.2 M_\odot$) dwarfs represent the lower-mass end of star formation through the process of collapsing molecular clouds (e.g., Luhman 2012), so studying planets around VLM dwarfs can test different planet formation theories in limiting conditions (e.g., Ida & Lin 2005; Boss 2006). Due to the intrinsic faintness of VLM dwarfs, the detection of planets around them is challenging for most of exoplanet detection methods such as the transit and the radial velocity methods. Although microlening planets comprise a minor fraction ($\sim 2.2\%$[1]) of all known planets, the technique plays an important role in probing planets orbiting VLM dwarfs because it does not rely on the light from the host stars but rather uses the light from a background source (Mao & Paczynski 1991; Gould & Loeb 1992). Among the 81 confirmed planets orbiting a VLM dwarf, 29 of them were found by the microlensing method. However, only seven such microlens planetary systems have unambiguous mass measurements: MOA-2007-BLG-192 (Bennett et al. 2008; Kubas et al. 2012), MOA-2010-BLG-073 (Street et al. 2013), OGLE-2012-BLG-0358 (Han et al. 2013), OGLE-2013-BLG-0102 (Jung et al. 2015), OGLE-2013-BLG-0341 (Jung et al. 2015), MOA-2013-BLG-605 (Sumi et al. 2016), OGLE-2016-BLG-1195 (Shvartzvald et al. 2017; Bond et al. 2017), while other systems require a Bayesian analysis based on a Galactic model to estimate the mass of the planetary systems.

The mass measurement of a microlens lens system is challenging. To measure the mass of a lens system, one needs two observables that yield mass-distance relations for the lens systems, i.e., any two of the angular Einstein radius $\theta_E$, the microlens parallax $\pi_E$ and the apparent brightness of the lens system. The detection of lens brightness can be achieved by high angular resolution imaging when the source and lens are resolved (e.g., Alcock et al. 2001; Kozłowski et al. 2007; Batista et al. 2015; Bennett et al. 2015; Bhattacharya et al. 2018; Vandorou et al. 2020; Bennett et al. 2020; Bhattacharya et al. 2020; Terry et al. 2021), but it is difficult for very faint VLM dwarfs. The measurements of $\theta_E$ and $\pi_E$ can yield the mass of a lensing object by (Gould 2000)

$$M_L = \frac{\theta_E}{\kappa \pi_E}, \qquad (1)$$

and its distance by

$$D_L = \frac{\text{AU}}{\pi_{\text{rel}} + \pi_S}, \qquad \pi_{\text{rel}} = \pi_E \theta_E, \qquad (2)$$

where $\kappa \equiv 4G/(c^2 \text{AU}) = 8.144$ mas/$M_\odot$, $\pi_S = \text{au}/D_S$ is the source parallax, $D_S$ is the source distance (Gould 1992, 2004) and $\pi_{\text{rel}}$ is the lens-source relative parallax. The measurements of angular Einstein radii $\theta_E$ are mainly via finite-source effects when the source crosses or approaches a caustic along the line of sight (Gould 1994; Witt & Mao 1994; Nemiroff & Wickramasinghe 1994), which are frequently detected in binary/planetary events because of their relatively large caustic structures. For the 29 microlens planet-VLM events, 24 of them have measurements of finite-source effects and thus the angular Einstein radius $\theta_E$. The microlens parallax $\pi_E$ can be measured by the annual parallax effect (Gould 1992), in which Earth's acceleration around the Sun introduces deviation from rectilinear motion to the lens-source relative motion. This method is generally feasible for events with long microlensing timescales $t_E \gtrsim$ year$/2\pi$ (e.g., Gaudi et al. 2008; Bennett et al. 2010) and/or events produced by nearby lenses (e.g., Jung et al. 2018). However, because the typical microlensing timescales for VLM events are $\lesssim 20$ days (see Equation 17 of Mao 2012), measurements of the annual parallax for planet-VLM events are challenging and only six of such events have a robust detection of annual parallax.

Microlens parallax $\pi_E$ can also be measured via "satellite microlens parallax", which is done by observing the same microlensing event from Earth and one or more well-separated ($\sim$ AU) satellite (Refsdal 1966; Gould 1994, 1995). The feasibility of satellite microlens parallax measurements has been demonstrated by the *Spitzer* satellite telescope (Dong et al. 2007; Udalski et al. 2015b; Yee et al. 2015a; Zhu et al. 2015; Calchi Novati et al. 2015a), the two-wheel *Kepler* satellite telescope (Zhu et al. 2017a; Zang et al. 2018; Poleski et al. 2019), the *Gaia* satellite (Wyrzykowski et al. 2020) and the joint observations of *Spitzer* and *Kepler* (Zhu et al. 2017c). Since 2014, the *Spitzer* satellite observed about 1100 microlensing events and yielded satellite parallax measurements for ten microlens planetary events: OGLE-2014-BLG-0124 (Udalski et al. 2015b; Beaulieu et al. 2018), OGLE-2015-BLG-0966 (Street et al. 2016), OGLE-2016-BLG-1067 (Calchi Novati et al. 2019), OGLE-2016-BLG-1190 (Ryu et al. 2018), OGLE-2016-BLG-1195 (Shvartzvald et al. 2017; Bond et al. 2017), OGLE-2017-BLG-0406 (Hirao et al. 2020), OGLE-2017-BLG-1140 (Calchi Novati et al. 2018), OGLE-2018-BLG-0596 (Jung et al. 2019), KMT-2018-BLG-0029 (Gould et al. 2020), Kojima-1 (Nucita et al. 2018; Fukui et al. 2019; Zang et al. 2020b). In particular, for OGLE-2016-BLG-1195, the *Spitzer* satellite parallax combined with the measurements of $\theta_E$ from ground-based data revealed that this planetary system is composed of an Earth-mass (∼ $1.4M_\oplus$) planet around a ∼ $0.078M_\odot$ ultracool dwarf with a lens distance of ∼ 3.9 kpc.

---

* E-mail: 3130102785@zju.edu.cn
[1] http://exoplanetarchive.ipac.caltech.edu as of 2020 October 10

Here we report the analysis of the second *Spitzer* planet orbiting a VLM dwarf, OGLE-2018-BLG-0799Lb. The paper is structured as follows. In Section 2, we describe the ground-based and *Spitzer* observations of the event. We then fit the ground-based data in Section 3 and fit the *Spitzer* satellite parallax in Section 4. We estimate the physical parameters of the planetary system in Section 5. Finally, implications of this work and discussion are given in Section 6 and 7, respectively.

## 2 OBSERVATIONS AND DATA REDUCTIONS

### 2.1 Ground-based Observations

OGLE-2018-BLG-0799 was first discovered by the Optical Gravitational Lensing Experiment (OGLE) collaboration (Udalski et al. 2015a) and alerted by the OGLE Early Warning System (Udalski et al. 1994; Udalski 2003) on 2018 May 13. The event was located at equatorial coordinates $(\alpha, \delta)_{J2000}$ = (18:13:50.16, −25:29:08.6), corresponding to Galactic coordinates $(\ell, b)$ = (6.12, −3.73). It therefore lies in OGLE field BLG545, with a cadence of 0.5-1 observations per night. These data were taken using the 1.3 m Warsaw Telescope equipped with a 1.4 deg² FOV mosaic CCD camera at the Las Campanas Observatory in Chile (Udalski et al. 2015a). About 50 days after OGLE's alert, the Microlensing Observations in Astrophysics (MOA; Bond et al. 2001) group also identified this event as MOA-2018-BLG-215. The MOA group conducts a high-cadence survey toward the Galactic bulge using its 1.8 m telescope equipped with a 2.2 deg² FOV camera at the Mt. John University Observatory in New Zealand (Sumi et al. 2016). The cadence of the MOA group for this event is $\Gamma \sim 1$ hr$^{-1}$ on average. This event was also observed by the Korea Microlensing Telescope Network (KMTNet) which consists of three 1.6 m telescopes equipped with 4 deg² FOV cameras at the Cerro Tololo Inter-American Observatory (CTIO) in Chile (KMTC), the South African Astronomical Observatory (SAAO) in South Africa (KMTS) and the Siding Spring Observatory (SSO) in Australia (KMTA) (Kim et al. 2016). It was recognized after the end of the 2018 season by KMTNet's event-finding algorithm as KMT-2018-BLG-1741 (Kim et al. 2018). The event lies in the KMTNet BLG31 field, which has a nominal cadence of $\Gamma$ = 0.4 hr$^{-1}$. However, from the start of the season through 25 June 2018, the cadence of KMTA and KMTS was altered to $\Gamma$ = 0.3 hr$^{-1}$. Thus, the second half of the light curve (including the planetary anomaly) has a higher cadence than the first half. The great majority of data were taken in the $I$ band for OGLE and KMTNet groups, and MOA-Red filter (which is similar to the sum of the standard Cousins $R$- and $I$-band filters) for the MOA group, with occasional observations made in the $V$ band for measurement of the source color.

On 2018 June 30 (UT 23:18), the *Spitzer* team realized that OGLE-2018-BLG-0799 was deviating from the point-lens point-source model based on the KMTNet observations taken in the previous 24 hours. At that point, they scheduled high-cadence follow-up observations by Las Cumbres Observatory (LCO) global network of telescopes and the 1.3m SMARTS telescope equipped with the optical/NIR ANDICAM camera at CTIO (CT13, DePoy et al. 2003). For this event, the LCO observations were taken by the 1m telescopes in CTIO and SSO, and the 0.4m telescopes in SSO, with SDSS-$i'$ filter. The majority of CT13 observations were taken in the $I$ band and $H$ band, with occasional observations in the $V$ band. The LCO 0.4m, CT13 $V$- and $H$-band data were excluded from the analysis due to excessive noise. In Table 1, we list details about the data used in the analysis.

### 2.2 *Spitzer* Observations

The goal of the *Spitzer* microlensing parallax program is to create an unbiased sample of microlensing events with well-measured parallax. In order to isolate the knowledge of the presence or absence of planets from influencing event selections, Yee et al. (2015b) developed protocols for selecting *Spitzer* targets. There are three ways an event may be selected for *Spitzer* observations. First, events that meet the specified objective criteria are selected as "objective" targets and *must* be observed with a pre-specified cadence. Second, events that do not meet these criteria can still be chosen as "subjective" targets *at any time for any reason*, but only data taken (or rather, made public) after this selection date can be used to calculate the planetary sensitivity of the events. The *Spitzer* team can publicly announce specified conditions for a candidate "subjective" target, and targets that obey the conditions are then automatically selected as a "subjective" target. "Subjective" selection is crucial because the "objective" criteria must be strictly defined so that all the "objective" targets have both high sensitivity to planets and a high likelihood of yielding a parallax measurement. In some cases, an event may never become objective but still be a good candidate. In addition, *Spitzer* observations that start a week or two earlier may improve the parallax measurement for an event that will meet the "objective" criteria later. Finally, events can be selected as "secret" targets without any announcement and become "subjectively selected" after the *Spitzer* team makes a public announcement.

Although OGLE-2018-BLG-0799 was recognized as a promising target early on, observations could not begin until July 9 due to Sun-angle constraints (the target is in the far western side of the bulge). It was announced as a candidate "subjective" *Spitzer* target on 2018 June 12, with a specified condition: if the $I$-band magnitude is brighter than 16.85 mag at HJD′ = 8301.5 (HJD′ = HJD − 2450000), the event would be "subjectively" selected. The event met this condition with $I$ = 16.36 at HJD′ = 8301.5. However, it did not meet the objective criteria because it had already peaked at $A_{\max} < 3$. Each *Spitzer* observation was composed of six dithered 30s exposures using the 3.6 $\mu$m channel ($L$−band) of the IRAC camera. *Spitzer* observed this event 31 times with a daily cadence in 2018. In order to test for systematic errors pointed out by Zhu et al. (2017b) and Koshimoto & Bennett (2020) (see Section 4.1 for details), OGLE-2018-BLG-0799 was reobserved at baseline five times over eight days near the beginning of the 2019 observing window.

### 2.3 Data Reduction

Data reductions of the OGLE, MOA, KMTNet and LCO datasets were conducted using custom implementations of the difference image analysis technique (Tomaney & Crotts 1996; Alard & Lupton 1998): Wozniak 2000 (OGLE), Bond et al. 2001 (MOA), Albrow et al. 2009 (KMTNet) and Bramich 2008 (LCO). The CT13 data were reduced using DoPHOT (Schechter et al. 1993). The *Spitzer* data were reduced using specially designed software for crowded-field photometry (Calchi Novati et al. 2015b). In addition, to measure the source color and construct the color-magnitude diagram (CMD), we conduct pyDIA photometry[2] for the KMTC data, which yields field-star photometry on the same system as the light curve.

**Table 1.** Data used in our analysis

| Collaboration | Site | Filter | Coverage (HJD') | $N_{\rm data}$ | Reduction Method |
|---|---|---|---|---|---|
| OGLE | | *I* | 7800 – 8398 | 158 | Wozniak (2000) |
| MOA | | Red | 8157 – 8392 | 486 | Bond et al. (2001) |
| KMTNet | SSO | *I* | 8171 – 8400 | 295 | pySIS (Albrow et al. 2009) |
| | CTIO | *I* | 8169 – 8412 | 435 | pySIS (Albrow et al. 2009) |
| | CTIO | *I* | 8169 – 8412 | 435 | pyDIA |
| | CTIO | *V* | 8176 – 8409 | 44 | pyDIA |
| | SAAO | *I* | 8172 – 8402 | 259 | pySIS (Albrow et al. 2009) |
| CT13 | | *I* | 8287 – 8329 | 49 | DoPHOT (Schechter et al. 1993) |
| LCO | SSO | *i* | 8302 – 8310 | 26 | DanDIA (Bramich 2008) |
| | CTIO | *i* | 8300 – 8311 | 20 | DanDIA (Bramich 2008) |
| *Spitzer* | | *L* | 8308 – 8690 | 36 | Calchi Novati et al. (2015b) |

[1] HJD' = HJD − 2450000

## 3 GROUND-BASED LIGHT CURVE ANALYSIS

Figure 1 shows the observed data together with the best-fit models. The ground-based light curve shows a bump (HJD' ∼ 8300) after the peak of an otherwise normal point-lens point-source light curve. The bump could be a binary-lensing (2L1S) anomaly or the second peak of a binary-source event (1L2S). Thus, we perform both 2L1S and 1L2S analysis in this section. Finally, in order to compare parallax constraints from ground-based data and *Spitzer* data to check against possible systematics in either data set, we fit the annual parallax effect in Section 3.3.

### 3.1 Static Binary-Lens Model

A "static" binary-lens model requires seven geometric parameters to calculate the magnification, $A(t)$. These include three point-lens parameters (Paczyński 1986)): the time of the maximum magnification, $t_0$, the minimum impact parameter, $u_0$, which is in units of the angular Einstein radius $\theta_{\rm E}$, and the Einstein radius crossing time, $t_{\rm E}$. There are four additional parameters: the angular radius of the source star, $\rho$, in units of $\theta_{\rm E}$; mass ratio of the binary, $q$; the projected separation, s, between the binary components normalized to $\theta_{\rm E}$; and the angle of source trajectory relative to the binary axis in the lens plane, $\alpha$. We use the advanced contour integration code (Bozza 2010; Bozza et al. 2018), VBBinaryLensing[3] to compute the binary-lens magnification $A(t)$. In addition, for each data set $i$, there are two linear parameters ($f_{{\rm S},i}, f_{{\rm B},i}$) representing the flux of the source star and any blended flux, respectively. Hence, the observed flux $f_i(t)$ is modeled as

$$f_i(t) = f_{{\rm S},i} A(t) + f_{{\rm B},i}. \tag{3}$$

[2] MichaelDAlbrow/pyDIA: Initial Release on Github, doi:10.5281/zenodo.268049
[3] http://www.fisica.unisa.it/GravitationAstrophysics/VBBinaryLensing.htm

In addition, we adopt a linear limb-darkening law to consider the brightness profile of the source star (An et al. 2002). According to the extinction-corrected source color and the color-temperature relation of Houdashelt et al. (2000), we estimate the effective temperature of the source to be $T_{\rm eff}$ ∼ 4900 K. Applying ATLAS models (Claret & Bloemen 2011), we obtain the linear limb-darkening coefficients $u_I = 0.56$ for the *I* band, $u_{i'} = 0.58$ for the SDSS-*i'* band, $u_R = 0.66$ for the *R* band (Claret & Bloemen 2011). For the MOA data, we adopt $\Gamma_{\rm MOA} = (\Gamma_I + \Gamma_R)/2 = 0.61$.

To search the parameter space of 2L1S models, we first carry out a sparse grid search on parameters ($\log s, \log q, \alpha, \log \rho$), with 21 values equally spaced between $−1 \leq \log s \leq 1$, $0° \leq \alpha < 360°$, 51 values equally spaced between $−5 \leq \log q \leq 0$ and 8 values equally spaced between $−3 \leq \log \rho \leq −1$, respectively. For each set of ($\log s, \log q, \alpha, \log \rho$), we fix $\log q, \log s, \log \rho$, with $t_0, u_0, t_{\rm E}, \alpha$ free. We find the minimum $\chi^2$ by Markov chain Monte Carlo (MCMC) $\chi^2$ minimization using the emcee ensemble sampler (Foreman-Mackey et al. 2013). The sparse grid search shows that the distinct minima are within $−0.2 \leq \log s \leq 0.3$ and $−4.5 \leq \log q \leq −1.5$. We then conduct a denser grid search, which consists of 51 values equally spaced between $−0.2 \leq \log s \leq 0.3$, and 31 values equally spaced between $−4.5 \leq \log q \leq −1.5$. As a result, we find three distinct minima and label them as models A, B and C in the Figure 2.

We then investigate the best-fit models by MCMC with all geometric parameters free. Finally, model A ($\log s, \log q) = (0.048 \pm 0.003, −2.58 \pm 0.02$) provides the best fit to the observed data, while model B ($\log s, \log q) = (0.151 \pm 0.002, −2.53 \pm 0.02$) and model C ($\log s, \log q) = (0.093 \pm 0.002, −3.46 \pm 0.02$) are disfavored by $\Delta\chi^2$ ∼ 68 and ∼ 61, respectively. In addition, finite-source effects of model A are marginally detected. The modeling only provides an upper limit on the source size normalized by the Einstein radius, $\rho < 0.026$ (3$\sigma$ level). The best-fit model has $\rho = 0.016$, but the data are also marginally consistent with a point-source model at $\Delta\chi^2 = 7$. Likewise for model B, the best-fit value of $\rho$ is 0.0002 but is consistent with zero in 1-$\sigma$ and has a 3-$\sigma$ upper limit of 0.010. For model C, finite source effects are measured to be $\rho = 0.0303 \pm 0.0009$. The best-fit parameters of the three models are given in Table 2, and the caustic geometries of the three models are shown in Figure 3.

We find that the MCMC does not jump from one model to the other in a normal run. To investigate the barriers between the three models and check for other potential degenerate models, we run a "hotter" MCMC by artificially inflating the error bars by a factor of 5.0. The upper panel of Figure 4 shows $\log q$ against the offset of the source trajectory from the planetary caustic center (Hwang et al. 2018a,b; Skowron et al. 2018)

$$\Delta \xi = u_0 \csc(\alpha) - (s - s^{-1}). \quad (4)$$

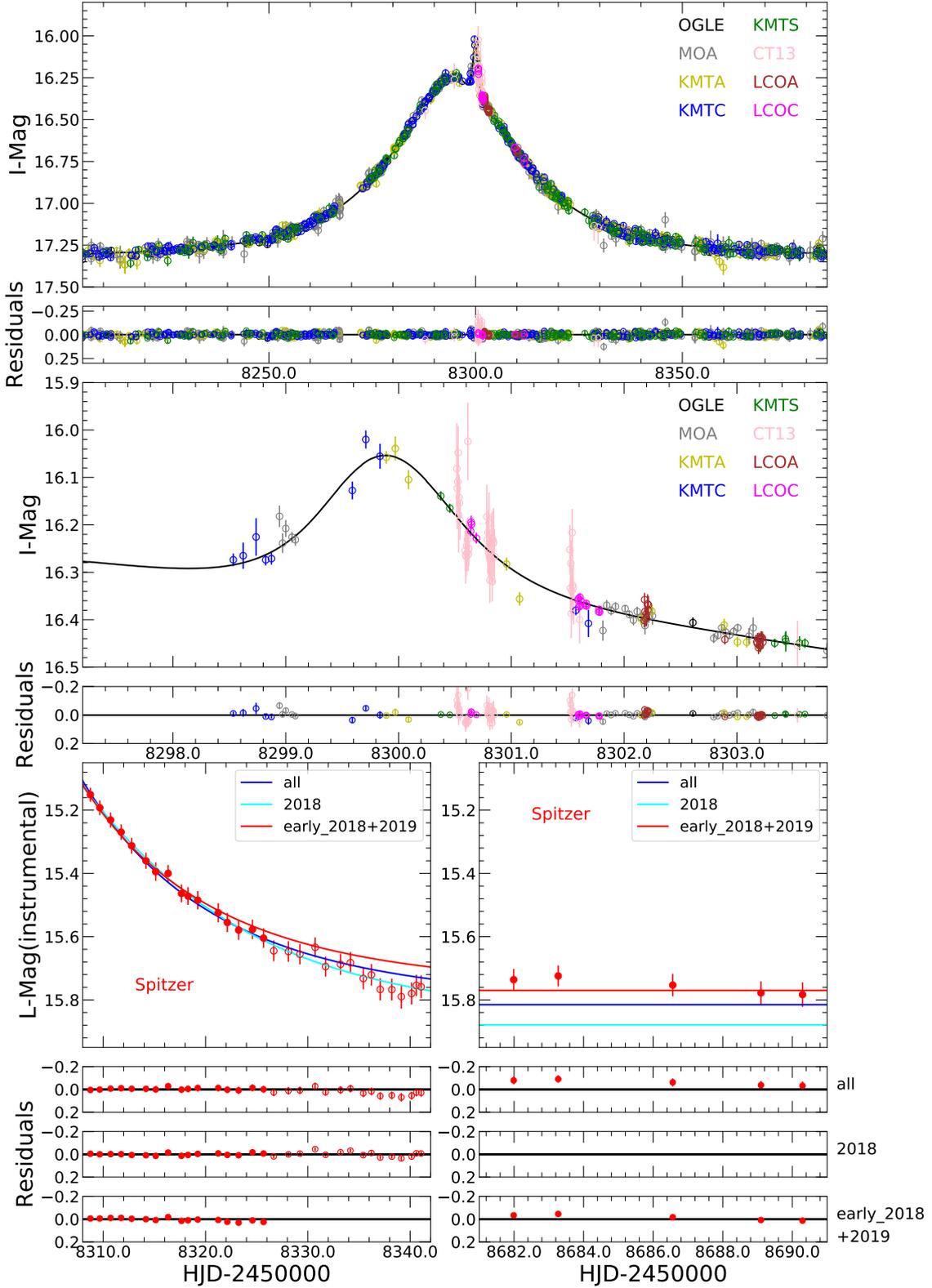

**Figure 1.** The observed data with the best-fit 2L1S model. The circles with different colors are observed data points for different data sets. The black solid line represents the best-fit model for the ground-based data. The middle panel shows a close-up of the planetary signal. The bottom panels show *Spitzer* observations with the residuals from the best-fit models. The *Spitzer* data in the "early_2018 + 2019" subset are shown as filled, red circles, while the "late 2018" data are shown as open circles. The best-fit models for each subset of the data "2018-only", "early_2018 + 2019", and "all" are shown as the cyan, red, and blue lines, respectively.

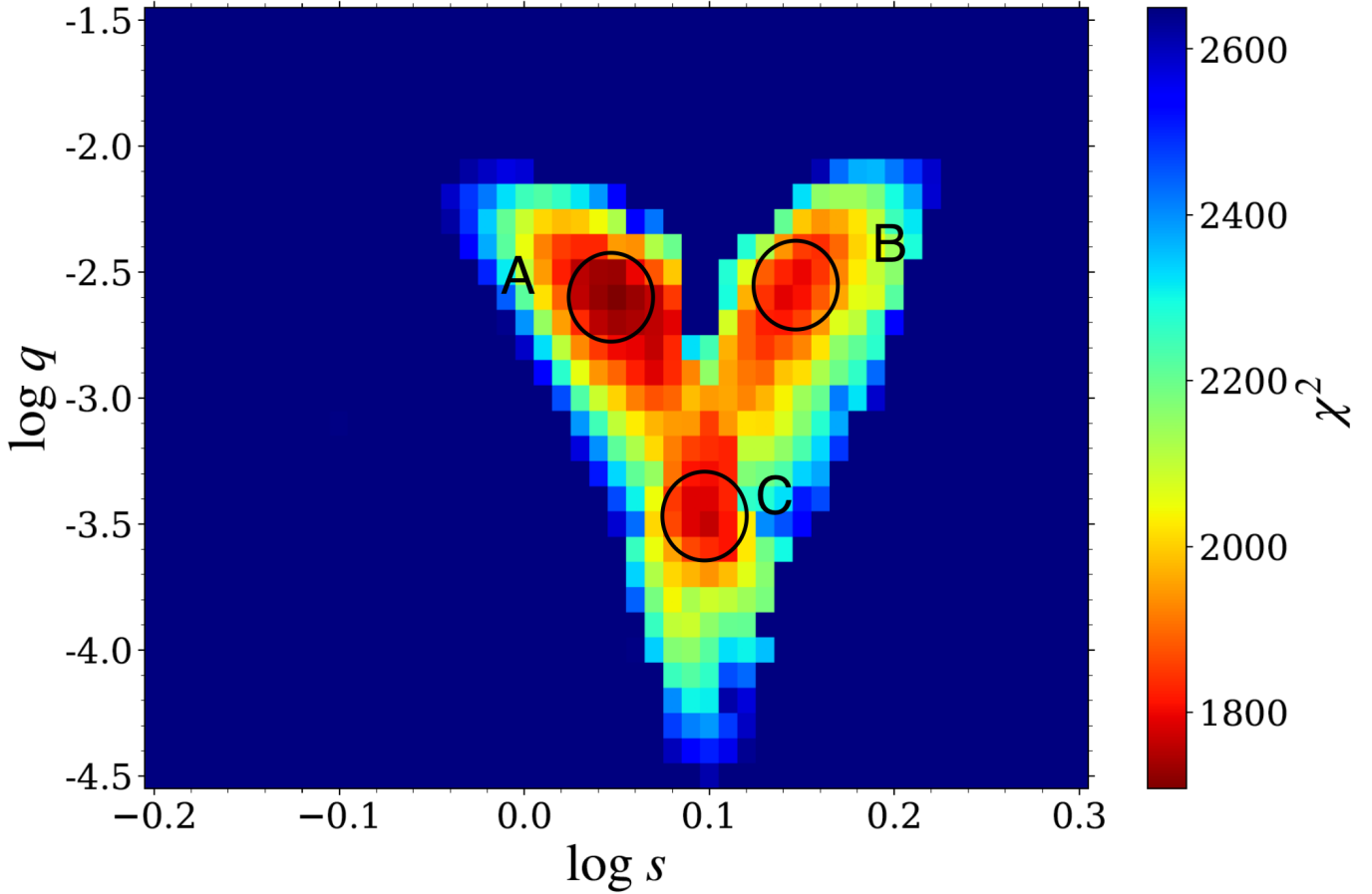

**Figure 2.** $\chi^2$ surface in the ($\log s$, $\log q$) plane from the grid search. The space is equally divided on a (51 × 31) grid with ranges of $-0.2 \leq \log s \leq 0.3$ and $-4.5 \leq \log q \leq -1.5$, respectively. The black circles labeled as A, B and C in the right panel represent three distinct minima.

We find that the barriers between the three models have $\Delta \chi^2 > 125$ and there is no obvious additional model. We also note that the topology of Model C is characterized by a large source that crosses a planetary caustic. A similar topology and light curve were found in the planetary event OGLE-2017-BLG-0173, except that the corresponding Model C is split into 2 local minima (see Figure 4 of Hwang et al. 2018b). Thus, we further investigate model C using a "hotter" MCMC with the error bars inflated by a factor of $\sqrt{5}$. The bottom panel of Figure 4 shows the result, in which we do not find any further degeneracy.

While OGLE-2018-BLG-0799 is qualitatively similar to OGLE-2017-BLG-0173, there are also notable differences in the two cases. Both have a single planetary perturbation dominated by finite source effects rather than a distinct caustic entrance and exit. The resulting $\chi^2$ surface in both cases has three minima, one in which the source passes directly over the planetary caustic (in the case of OGLE-2017-BLG-0173, this minimum is bimodal) and two in which the source passes to one side or the other of the planetary caustic. However, in the case of OGLE-2017-BLG-0173, in the solution with the source passing directly over the caustic, the source is much larger than the caustic, whereas in the solutions in which the source passes to one side or the other, the source size is comparable to the size of the caustic. By contrast, in the present case, when the source passes directly over the caustic, it is comparable in size to the caustic (see Figure 3) but when it passes to one side of the other, it does not cross the caustic and there is only an upper limit on the source size. In addition, in OGLE-2017-BLG-0173, the degeneracies between the solutions cannot be definitely resolved, whereas in the present case, the degeneracy between the three solutions is clearly resolved by $\chi^2$.

In Figure 5, we show the residuals for the three models and draw the cumulative $\Delta \chi^2$ distribution of models B and C relative to model A over the anomaly region. We find that most of the $\chi^2$ differences are from the short-lived bump, and models B and C cannot well fit the data over the anomaly region. We also check whether the $\Delta \chi^2$ can be decreased by considering the parallax effect, but all of models in Sections 3.3 and 4 have $\Delta \chi^2 > 60$ for models B and C. Thus, we exclude models B and C.

### 3.2 Binary-Source Model

Gaudi (1998) first pointed out that a binary-source event can also cause a smooth, short-lived, low-amplitude bump if the second source is much fainter and passes closer to the lens, which is similar to planet-induced anomalies. The total magnification of a 1L2S event is the superposition of two point-lens events,

$$A_\lambda = \frac{A_1 f_{1,\lambda} + A_2 f_{2,\lambda}}{f_{1,\lambda} + f_{2,\lambda}} = \frac{A_1 + q_{f,\lambda} A_2}{1 + q_{f,\lambda}}, \quad (5)$$

$$q_{f,\lambda} = \frac{f_{2,\lambda}}{f_{1,\lambda}}, \quad (6)$$

where $f_{i,\lambda}$ ($i = 1, 2$) is the flux at wavelength $\lambda$ of each source and $A_\lambda$ is total magnification (Hwang et al. 2013). The best-fit 1L2S model is disfavored by $\Delta\chi^2 \sim 157$ compared to the 2L1S model A (see Table 2 for the parameters). In Figure 5, we find that the $\chi^2$ difference to the 2L1S model A is mainly from the short-lived bump and the 1L2S model fails to fit the observed data. Thus, we exclude the 1L2S model.

**Table 2.** Best-fit models and their 68% uncertainty ranges from MCMC for ground-only data

| Models | 2L1S | | | 1L2S |
|---|---|---|---|---|
| | A | B | C | |
| $\chi^2/dof$ | 1704.8/1705 | 1771.5/1705 | 1765.8/1705 | 1861.7/1704 |
| $t_{0,1}$ (HJD$'$) | 8295.15 ± 0.02 | 8295.26 ± 0.02 | 8295.10 ± 0.02 | 8294.87 ± 0.02 |
| $t_{0,2}$ (HJD$'$) | ... | ... | ... | 8300.04 ± 0.04 |
| $u_{0,1}$ | 0.403 ± 0.008 | 0.409 ± 0.009 | 0.397 ± 0.008 | 0.620 ± 0.041 |
| $u_{0,2}$ | ... | ... | ... | 0.000 ± 0.002 |
| $t_E$ (days) | 28.2 ± 0.4 | 27.8 ± 0.4 | 28.7 ± 0.4 | 24.6 ± 1.0 |
| $s$ | 1.116 ± 0.008 | 1.415 ± 0.007 | 1.239 ± 0.006 | ... |
| $q (10^{-3})$ | 2.62 ± 0.14 | 2.98 ± 0.15 | 0.345 ± 0.015 | ... |
| $\alpha$ (rad) | 1.170 ± 0.003 | 1.178 ± 0.003 | 1.174 ± 0.003 | ... |
| $\rho_1$ | < 0.026 | < 0.010 | 0.0303 ± 0.0009 | 0.606 ± 0.050 |
| $\rho_2$ | ... | ... | ... | 0.020 ± 0.002 |
| $q_{f,I}$ | ... | ... | ... | 0.0049 ± 0.0003 |
| $f_{S,OGLE}$ | 1.794 ± 0.047 | 1.850 ± 0.054 | 1.727 ± 0.049 | 2.372 ± 0.176 |
| $f_{B,OGLE}$ | −0.032 ± 0.046 | −0.086 ± 0.052 | 0.034 ± 0.048 | −0.612 ± 0.175 |

[1] The values of $\rho_1$ are their $3\sigma$ upper limits. All fluxes are on an 18th magnitude scale, e.g., $I_S = 18 - 2.5\log(f_S)$.

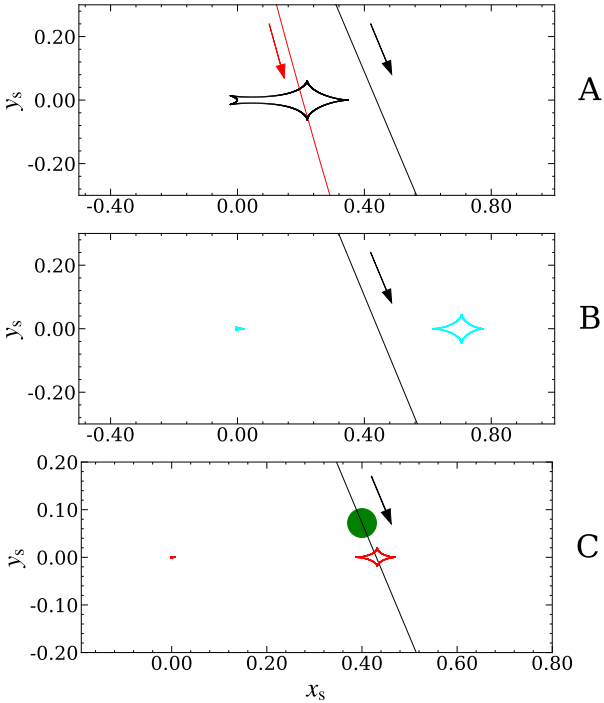

**Figure 3.** Caustic geometries of the three static 2L1S models. The caustics are color-coded to match the light curves in the Figure 5. The axes are in units of the Einstein radius $\theta_E$. In each panel, the black solid line is the source trajectory seen from the ground, and the arrow indicates the direction of the source motion. In the top panel, the red solid line is the source trajectory seen from the *Spitzer* satellite. Because finite source effects are measured for model C, the radius of the green circle in the bottom panel represents the source radius $\rho = 0.0303$. Models A and B only have weak constraints on $\rho$ (see Section 3.1), so their source radii are not shown.

### 3.3 Ground-Based Parallax

We fit the annual parallax effect by introducing two additional parameters $\pi_{E,N}$ and $\pi_{E,E}$, the North and East components of $\vec{\pi}_E$ in equatorial coordinates (Gould 2004). Because the annual parallax effect can be correlated with the effects of lens orbital motion, we also introduce two parameters $(ds/dt, d\alpha/dt)$, the instantaneous changes in the separation and orientation of the two components defined at $t_0$, for linearized orbital motion. We find that the orbit parameters are relatively poorly constrained, and we therefore restrict the MCMC trials to $\beta < 0.8$, where $\beta$ is the ratio of projected kinetic to potential energy (Dong et al. 2009)

$$\beta \equiv \left|\frac{KE_\perp}{PE_\perp}\right| = \frac{\kappa M_\odot yr^2}{8\pi^2} \frac{\pi_E}{\theta_E} \gamma^2 \left(\frac{s}{\pi_E + \pi_S/\theta_E}\right)^3; \quad \vec{\gamma} \equiv \left(\frac{ds/dt}{s}, \frac{d\alpha}{dt}\right), \quad (7)$$

and we adopt $\pi_S = 0.13$ mas for the source parallax, $\theta_* = 2.75$ $\mu$as from Section 5.1 (and thus, $\theta_E = \theta_*/\rho$). We also fit $u_0 > 0$ and $u_0 < 0$ models to consider the "ecliptic degeneracy" (Jiang et al. 2004; Poindexter et al. 2005). See the top panels of Figure 6 for the error contours of annual parallax. For both $u_0 > 0$ and $u_0 < 0$ models, we find that the $\chi^2$ improvement relative to the static model is only 1.5 and $\pi_{E,E}$ has a best-fit value of $\sim -0.3$ with an 1-$\sigma$ error of 0.27. For the $u_0 > 0$ model, $\pi_{E,N}$ has an 1-$\sigma$ error of 0.15, while $\pi_{E,N}$ is only broadly constrained for the $u_0 < 0$ model. The effects of lens orbital motion is not detectable ($\Delta \chi^2 = 0.2$) and not significantly correlated with $\vec{\pi}_E$, so we eliminate the lens orbital motion from the fit[4].

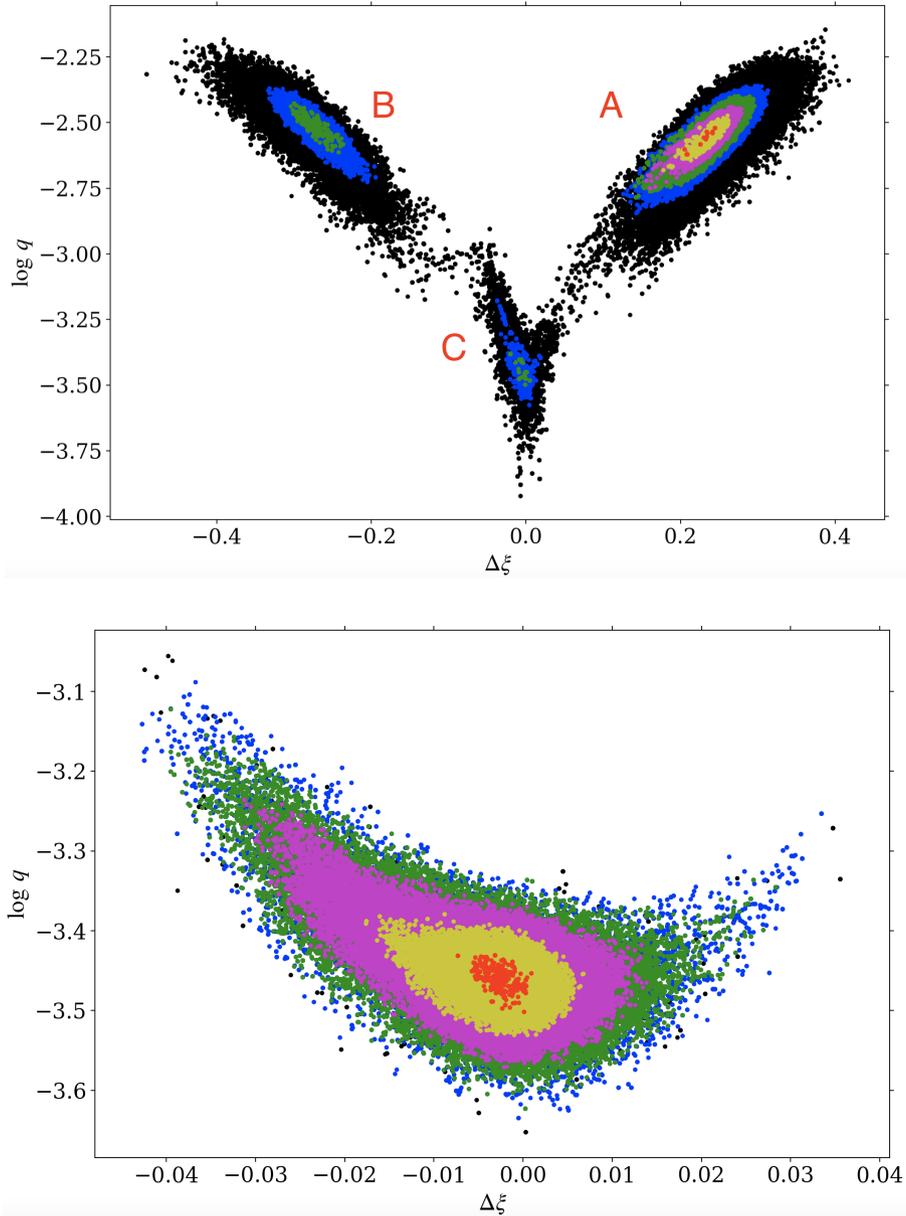

**Figure 4.** Scatter plot of $\Delta \xi$ vs. $\log q$ from "hotter" MCMC chains, where $\Delta \xi = u_0 \csc(\alpha) - (s - s^{-1})$ is the offset of the center of the source from the center of the caustic at the moment that the source crosses the binary axis. Upper panel: the result is derived by inflating the error bars by a factor of 5.0, and then multiplying the resulting $\chi^2$ by 25 for the plot. Lower panel: the result is derived by inflating the error bars by a factor of $\sqrt{5}$, and then multiplying the resulting $\chi^2$ by 5 for the plot. Note that the best-fit solution shown in the upper panel is preferred over that shown in the lower panel by $\Delta \chi^2 = 61$. The purpose of the lower panel is to check whether the model C has a bimodal minimum similar to the corresponding model of OGLE-2017-BLG-0173 (Hwang et al. 2018b). In each panel, the initial parameters of the MCMC chain are the Model C shown in Table 2. Red, yellow, magenta, green, blue and black colors represent $\Delta \chi^2 < 5 \times (1, 4, 9, 16, 25, \infty)$.

## 4 PARALLAX ANALYSIS INCLUDING *Spitzer* DATA

Simultaneous observations from two widely separated observers can result in two different observed light curves (Refsdal 1966), which yields the measurement of the microlens parallax (see Figure 1 of Gould 1994),

$$\vec{\pi}_E = \frac{au}{D_\perp}(\Delta\tau, \Delta\beta),  \quad (8)$$

with

$$\Delta\tau \equiv \frac{t_{0,Spitzer} - t_{0,\oplus}}{t_E}; \quad \Delta\beta \equiv \pm u_{0,Spitzer} - \pm u_{0,\oplus}, \quad (9)$$

where $D_\perp$ is the projected separation between the *Spitzer* satellite and Earth at the time of the event. In addition, we include a *VIL* color-color constraint on the *Spitzer* source flux $f_{s,Spitzer}$ (e.g., Shin et al. 2018), which adds a $\chi^2_{\rm penalty}$ into the total $\chi^2$,

$$\chi^2_{\rm penalty} = \frac{[(I-L)_S - (I-L)_{\rm fix}]^2}{\sigma^2_{cc}}, \quad (10)$$

where $(I-L)_S$ is the source color from the modeling, $(I-L)_{\rm fix}$ is the color constraint, and $\sigma_{cc}$ is the uncertainty of the color constraint. To derive the color-color constraint of the *Spitzer* source flux, we extract *Spitzer* and KMTC photometry for the stars within the range $1.8 < (V-I)_{\rm KMT} < 2.5$, which have color close to the source star. We obtain the color-color relation

$$I_{\rm KMT} - L_{Spitzer} = 1.74 + [1.38(V-I)_{\rm KMT} - 2.08]. \quad (11)$$

---

[4] We also check whether the constraint of lens orbital motion could be improved by including *Spitzer* data, but the constraint is still weak ($\Delta \chi^2 < 1$).

Table 3. Parallax models for the solution A for ground-only data

|  | $u_0 > 0$ | | $u_0 < 0$ | |
| --- | --- | --- | --- | --- |
| Models | Parallax | Parallax + Orbit | Parallax | Parallax + Orbit |
| $\chi^2/dof$ | 1703.3/1703 | 1703.1/1701 | 1703.3/1703 | 1703.1/1701 |
| $t_0$ (HJD′) | 8295.16 ± 0.02 | 8295.15 ± 0.03 | 8295.15 ± 0.02 | 8295.16 ± 0.04 |
| $u_0$ | 0.413 ± 0.012 | 0.419 ± 0.015 | −0.417 ± 0.013 | −0.418 ± 0.014 |
| $t_E$ (days) | 27.7 ± 0.6 | 27.5 ± 0.7 | 27.5 ± 0.6 | 27.5 ± 0.7 |
| s | 1.123 ± 0.012 | 1.128 ± 0.024 | 1.127 ± 0.012 | 1.128 ± 0.025 |
| $q(10^{-3})$ | 2.61 ± 0.16 | 2.52 ± 0.49 | 2.58 ± 0.17 | 2.61 ± 0.77 |
| $\alpha$ (rad) | 1.173 ± 0.003 | 1.175 ± 0.004 | −1.175 ± 0.003 | −1.181 ± 0.004 |
| $\rho$ | < 0.024 | < 0.025 | < 0.025 | < 0.025 |
| $\pi_{E,N}$ | 0.076 ± 0.154 | 0.097 ± 0.164 | −0.366 ± 0.592 | −0.798 ± 0.720 |
| $\pi_{E,E}$ | −0.317 ± 0.272 | −0.425 ± 0.307 | −0.354 ± 0.275 | −0.399 ± 0.317 |
| $ds/dt$ (yr$^{-1}$) | ... | 0.058 ± 0.943 | ... | 0.546 ± 0.981 |
| $d\alpha/dt$ (yr$^{-1}$) | ... | 0.103 ± 2.575 | ... | 0.340 ± 4.790 |
| $f_{S,OGLE}$ | 1.886 ± 0.080 | 1.897 ± 0.097 | 1.905 ± 0.082 | 1.894 ± 0.094 |
| $f_{B,OGLE}$ | −0.128 ± 0.081 | −0.139 ± 0.099 | −0.146 ± 0.084 | −0.135 ± 0.096 |

[1] The values of $\rho$ are their $3\sigma$ upper limits. All fluxes are on an 18th magnitude scale, e.g., $I_S = 18 - 2.5\log(f_S)$.

In Section 5, we find $(V-I)_{\rm KMT} = 2.035 \pm 0.018$. Hence,

$$(I_{\rm KMT} - L_{Spitzer})_{\rm fix} = 1.678 \pm 0.026. \quad (12)$$

### 4.1 *Spitzer* Systematics Investigation and *Spitzer*-"ONLY" Parallax

Poleski et al. (2016) were the first to discuss correlated noise in the *Spitzer* data, and Zhu et al. (2017b) found that the *Spitzer* data of six events have prominent deviations (see their Figure 6 for an example), which is likely due to systematics in the *Spitzer* data. Koshimoto & Bennett (2020) conducted a quantitative statistical test to the 50-event statistical sample of Zhu et al. (2017b) and found a conflict between the *Spitzer* microlensing parallax measurements and the predictions from Galactic models.

In order to check for the impact of systematic errors on the measured parallax, all known *Spitzer* planetary events from 2014–2018 were reobserved for about a week at baseline during the (final) 2019 season. Gould et al. (2020) first tested the *Spitzer* systematics using these baseline data for the event KMT-2018-BLG-0029. They divided the 2018 *Spitzer* data into two parts ("first-half-2018" and "second-half-2018") and combined each part with the 2019 data, as well as discussing the results from 2018 data alone. They conclude that the "second-half-2018" *Spitzer* data show clear residuals that are correlated in time, so should be excluded. Subsequently, in the event OGLE-2017-BLG-0406, Hirao et al. (2020) found a similar effect. They analyzed the *Spitzer* parallax with 2019 baseline data and found evidence of systematic errors in the last six *Spitzer* data points from 2017. They repeated the analysis with and without those six points and show that they only affect the *Spitzer* parallax measurements at less than $1\sigma$. Furthermore, they discuss the effects on resulting parallax contours in the context of the Gould (2019) osculating circles formalism. In these two cases, the subsets of *Spitzer* data with clear systematic errors are in the second half of the *Spitzer* observing season. Because the observing conditions (such as the rotation of the spacecraft or angle with the Sun) change over the course of a given season, the first half of the data from each season has more similar observing conditions to the 2019 baseline data, which were taken at the beginning of the 2019 season.

We follow the established procedure of Gould et al. (2020) and Hirao et al. (2020) to investigate the potential *Spitzer* systematics for OGLE-2018-BLG-0799. We perform this analysis on both the full *Spitzer* dataset ("all") and two subsets of the data ("2018" and "early_2018 + 2019").

For the "2018" subset, we use only data taken in 2018, and we exclude the 2019 baseline observations. This dataset matches what would be used for a "normal" *Spitzer* event. Thus, for the statistical

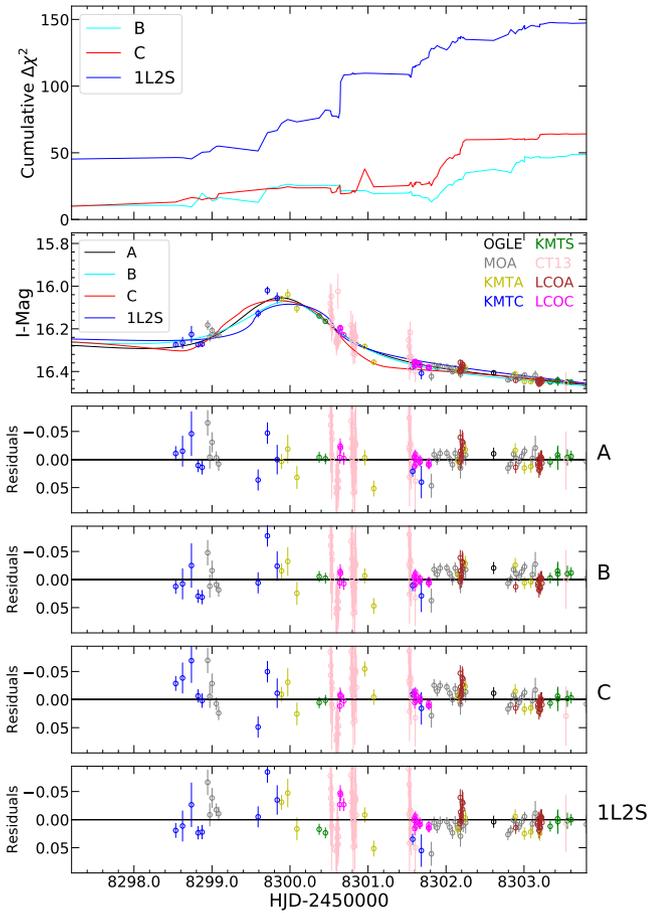

**Figure 5.** The upper panel shows the cumulative distribution of $\chi^2$ differences for 2L1S and 1L2S models compared to the 2L1S Model A ($\Delta\chi^2 = \chi^2_{\text{model}} - \chi^2_A$) over the anomaly region. The second panel shows a close-up of the anomaly region, in which the lines with different colors represent the different models. The residuals for each model are shown separately in the bottom four panels.

sample, membership in that sample should be determined solely on the basis of these data.

Following the method of Gould et al. (2020), we also fit just the first half (HJD' < 8326) of the 2018 *Spitzer* data combined with the 2019 baseline data (i.e., the "early_2018 + 2019" subset). As discussed in Gould et al. (2020), because the 2019 data were taken near the beginning of the 2019 observing window, the observing conditions are more similar at the beginning of the 2018 observing window.

In order to isolate the satellite parallax signal due to *Spitzer*'s separation from Earth, we first fit for the *Spitzer*-"ONLY" parallax (Jung et al. 2019). We fix ($t_0, u_0, t_E, s, q, \alpha, \rho$) along with the KMTC source flux as the best-fit parameters for the ground-based static models (shown in Table 2), and then fit ($\pi_{E,N}, \pi_{E,E}, f_{S,Spitzer}, f_{B,Spitzer}$) with the derived $VIL$ color-color constraint (Equation 12). We repeat the analysis for both the $u_0 > 0$ and $u_0 < 0$ solutions.

The error contours for the parallax vector derived from this analysis are shown in the middle panels of Figure 6. Although the $\chi^2$ contours have a similar arc-like form in all three cases, the measured value of the parallax and the 3-$\sigma$ uncertainties are disjoint (or close to disjoint) for the "2018" subset of the data as compared to the subsets including the 2019 baseline data. Because Gould et al. (2020) and Hirao et al. (2020) have shown that results of the subset with similar observing conditions are the least likely to be affected by the *Spitzer* systematics, we adopt the results for the "early_2018 + 2019" subset as our fiducial values. Indeed, adding the second half of the 2018 data, i.e., using all the data, stretches the contours toward the 2018-only result rather than refining them, indicating some effects from systematics. We discuss the implications of these discrepancies in detail in Sections 5 and 6.

### 4.2 Full Parallax Models

We finally fit the parallax combining ground-based and *Spitzer* data together. The resulting parallax contours are shown in the bottom panels of Figure 6, and the resulting parameters are shown in Table 4. There is some tension between the annual parallax constrained by ground-based data alone and the parallax measured from the *Spitzer* light curve. In particular, the annual parallax prefers a negative value of $\pi_{E,E}$ ($\sim -0.3$), whereas the *Spitzer*-"ONLY" parallax prefers a positive value of $\pi_{E,E}$ ($\sim 0.1$) when the 2019 baseline data are included. The tension with the annual parallax suggests the constraint is driven by some systematics in the ground-based data or stellar variability of the source star. However, we were unable to definitively identify the cause of the discrepancy or source of the systematics. Regardless, because the constraints from the annual parallax are broad, when the two effects are combined, the final result is dominated by the *Spitzer* parallax.

## 5 PHYSICAL PROPERTIES

Our physical interpretation of the lens is substantially different with and without the 2019 *Spitzer* baseline data. To simplify the discussion and show how the problem derives primarily from the parallax measurement itself, we begin in Section 5.1 by estimating the angular source radius $\theta_*$ and the angular Einstein radius $\theta_E$. Then, in Section 5.2, we examine the constraints on the lens mass $M_L$ and distance $D_L$ derived directly from $\theta_E$ and $|\vec{\pi}_E|$. Finally, in Section 5.3, we carry out a full Bayesian analysis to derive the properties of the lens weighted by a Galactic model.

### 5.1 Color Magnitude Diagram

The angular Einstein radius $\theta_E = \theta_*/\rho$. Thus, we estimate $\theta_*$ using a color magnitude diagram (CMD) analysis (Yoo et al. 2004). We construct a KMTC $V - I$ versus $I$ CMD using stars within a 120″ square centered on the source position (see Figure 7). We measure the centroid of the red clump $(V-I, I)_{cl} = (2.09 \pm 0.01, 15.83 \pm 0.04)$. We determine the source color by regression of $V$ versus $I$ flux as the source magnification changes, and find the source position $(V-I, I)_S = (2.035 \pm 0.018, 17.46 \pm 0.03)$. From Bensby et al. (2013); Nataf et al. (2013), the intrinsic color and de-reddened brightness of the red clump are $(V-I, I)_{cl,0} = (1.06, 14.27)$. Assuming the source suffers from the same dust extinction as the red clump, the intrinsic color and de-reddened magnitude of the source are $(V-I, I)_{S,0} = (1.00 \pm 0.03, 15.90 \pm 0.05)$. Using the color/surface-brightness relation of Adams et al. (2018), we obtain

$$\theta_* = 2.75 \pm 0.20 \, \mu\text{as}. \qquad (13)$$

By itself, the $3\sigma$ upper limit from $\rho$ alone yields a lower limit on the angular Einstein radius of $\theta_E > 0.106$ mas. However, there is, in fact a preferred value of $\rho$ from the fitting. Thus, combining the probability distribution function for $\rho$ from the full parallax models with $\theta_*$, yields a probability distribution of $\theta_E$, which is shown in Figure 8.

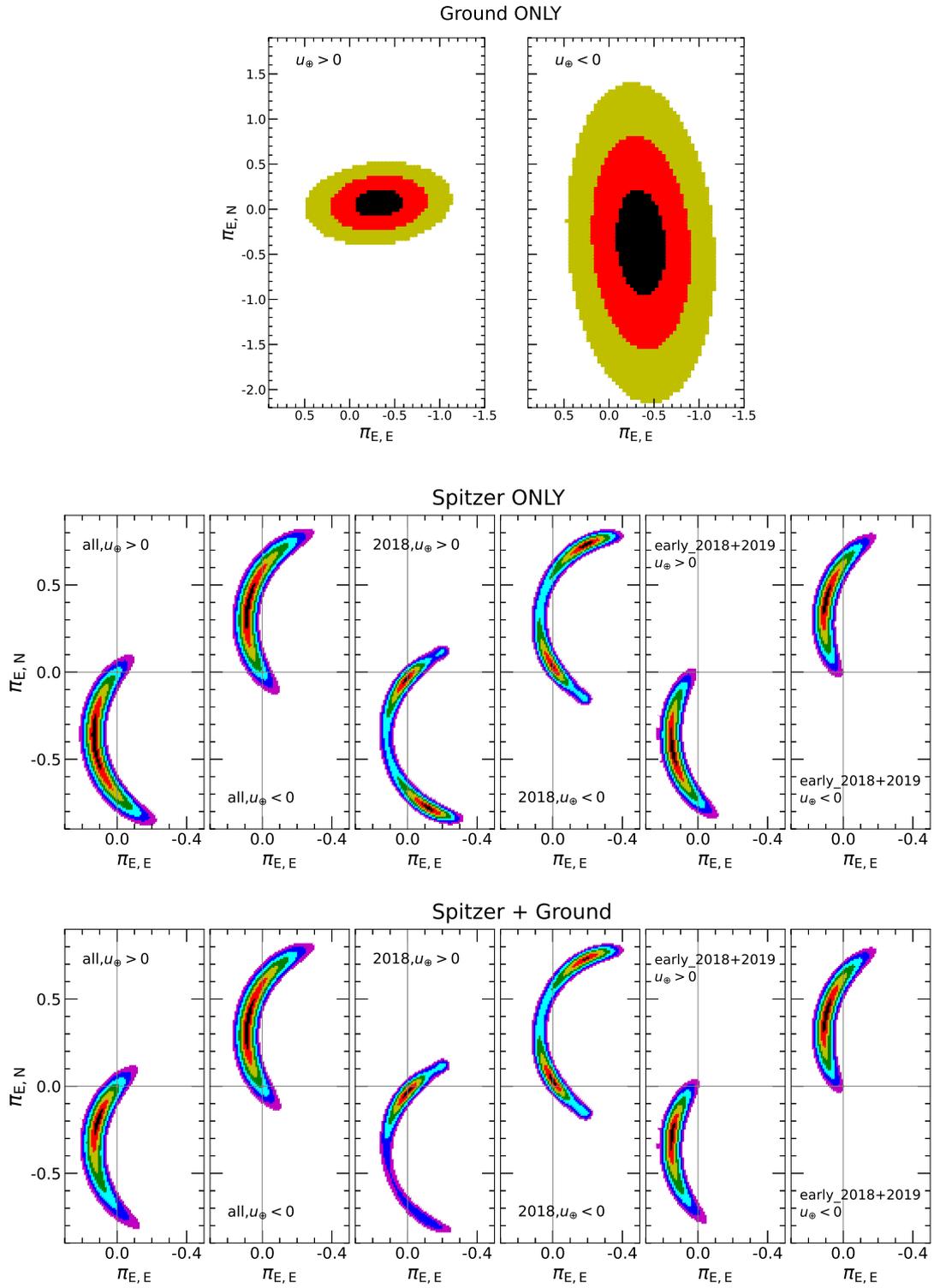

**Figure 6.** Parallax constraints from Ground-ONLY (top panels), *Spitzer*-"ONLY" (middle panels) and Ground+*Spitzer* (bottom panels) parallax analysis. Colors (black, red, yellow, green, cyan, blue, magenta) indicate number of $\sigma$ from the minimum (1, 2, 3, 4, 5, 6, 7). From left to right for the middle and bottom panels, contours are shown for the full *Spitzer* data set, the 2018 data alone, and the first half of 2018 + the 2019 data (referred to as the "early_2018 + 2019" subset in the text).

**Table 4.** Best-fit models and their 68% uncertainty ranges from MCMC for full parallax models

| Models | "all" | | "2018" | | "early_2018 + 2019" | |
|---|---|---|---|---|---|---|
| Parameters | $u_{0,\oplus} > 0$ | $u_{0,\oplus} < 0$ | $u_{0,\oplus} > 0$ | $u_{0,\oplus} < 0$ | $u_{0,\oplus} > 0$ | $u_{0,\oplus} < 0$ |
| $\chi^2_{\rm total}/dof$ | 1744.5/1737 | 1742.1/1737 | 1734.1/1732 | 1734.3/1732 | 1729.9/1722 | 1726.5/1722 |
| $t_0$ (HJD$'$) | 8295.13 ± 0.02 | 8295.13 ± 0.02 | 8295.13 ± 0.02 | 8295.15 ± 0.02 | 8295.13 ± 0.02 | 8295.13 ± 0.02 |
| $u_{0,\oplus}$ | 0.400 ± 0.009 | −0.402 ± 0.009 | 0.401 ± 0.009 | −0.404 ± 0.010 | 0.399 ± 0.009 | −0.403 ± 0.009 |
| $t_E$ (days) | 28.4 ± 0.4 | 28.2 ± 0.4 | 28.3 ± 0.4 | 28.1 ± 0.4 | 28.4 ± 0.4 | 28.2 ± 0.4 |
| $s$ | 1.111 ± 0.009 | 1.114 ± 0.009 | 1.115 ± 0.009 | 1.116 ± 0.009 | 1.112 ± 0.009 | 1.117 ± 0.009 |
| $q(10^{-3})$ | 2.70 ± 0.16 | 2.70 ± 0.16 | 2.62 ± 0.16 | 2.65 ± 0.16 | 2.65 ± 0.16 | 2.64 ± 0.16 |
| $\alpha$ (rad) | 1.168 ± 0.003 | 1.169 ± 0.003 | 1.167 ± 0.003 | −1.170 ± 0.003 | 1.167 ± 0.003 | −1.169 ± 0.003 |
| $\rho$ | < 0.026 | < 0.026 | < 0.026 | < 0.026 | < 0.026 | < 0.026 |
| $\pi_{E,N}$ | −0.218 ± 0.082 | 0.373 ± 0.126 | −0.037 ± 0.034 | 0.047 ± 0.043 | −0.301 ± 0.077 | 0.410 ± 0.090 |
| $\pi_{E,E}$ | 0.121 ± 0.021 | 0.083 ± 0.030 | 0.006 ± 0.027 | −0.020 ± 0.025 | 0.152 ± 0.012 | 0.109 ± 0.023 |
| $f_{S,OGLE}$ | 1.771 ± 0.053 | 1.790 ± 0.053 | 1.781 ± 0.053 | 1.802 ± 0.053 | 1.769 ± 0.052 | 1.747 ± 0.053 |
| $f_{B,OGLE}$ | −0.007 ± 0.052 | −0.025 ± 0.052 | −0.018 ± 0.052 | −0.039 ± 0.052 | −0.004 ± 0.051 | 0.017 ± 0.052 |
| $f_{S,Spitzer}$ | 7.633 ± 0.296 | 7.815 ± 0.297 | 7.818 ± 0.293 | 7.928 ± 0.294 | 7.671 ± 0.289 | 7.523 ± 0.293 |
| $f_{B,Spitzer}$ | −0.162 ± 0.294 | −0.334 ± 0.294 | −0.783 ± 0.305 | −0.872 ± 0.302 | 0.125 ± 0.287 | 0.267 ± 0.290 |
| $\chi^2_{\rm penalty}$ | 0.265 | 0.001 | 0.070 | 0.122 | 0.065 | 0.053 |

[1] From left to right, parameters are shown for the full *Spitzer* dataset ("all"), the 2018 data alone ("2018"), and the first half of 2018 + the 2019 data ("early_2018 + 2019"). The values of $\rho$ are their $3\sigma$ upper limits. All fluxes are on an 18th magnitude scale, e.g., $L_{S,Spitzer} = 18 − 2.5\log(f_{S,Spitzer})$.

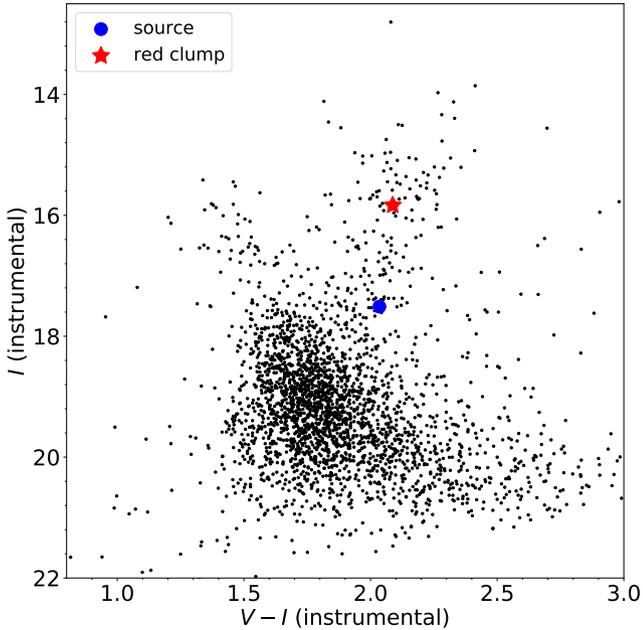

**Figure 7.** Instrumental color-magnitude diagram of a 120″ square centered on OGLE-2018-BLG-0799 using KMTC data. The red asterisk and blue dot represent the centroid of the red clump and the position of the microlens source, respectively.

### 5.2 Approximate

From the microlensing light curve, the ground-based data give a constraint on $\theta_E$ and the *Spitzer* data give a measurement of $|\vec{\pi}_E|$. Each of $\theta_E$ and $|\vec{\pi}_E|$ yields a mass-distance relationship (Equations 1 and 2) as shown in the left-hand panels of Figure 9. For the $|\vec{\pi}_E|$ constraint, we used the minimum $\chi^2$ for a given radius $\pi_E$ from the contours shown in Figure 6. For simplicity, we focus this discussion on the $u_0 < 0$ solution and the parallaxes derived from the "early_2018 + 2019" and "2018" subsets of the *Spitzer* data (the similarity in the parallax contours means that the $u_0 > 0$ solution and/or full *Spitzer* dataset yield qualitatively similar results). The $\theta_E$ relation is the same in all cases. The 1, 2, and 3-$\sigma$ limits for this relation are derived from the probability distribution shown in Figure 8.

The "early_2018 + 2019" case yields the simple intersection of two relations, but the "2018" case yields bimodal values for the parallax and hence, a pair of intersections with the $\theta_E$ constraint. However, we can also take into account the fact that more distant lenses are more likely, because the volume of stars is larger at larger distances for fixed $\theta_E$. Thus, we sum the $\chi^2$s from the two constraints and weight by a factor of $D_L^2$ to produce 1-, 2-, and 3-$\sigma$ contours for the lens mass and distance (right-hand panels of Figure 9). This downweights the smaller $D_L$ minimum (corresponding to the parallax minimum with larger $|\vec{\pi}_E|$) in the "2018" case. Finally, we find for the "early_2018 + 2019" case that the lens primary is a very low-mass object. By contrast, the 2018 data alone suggest that the lens is likely to be a $K$- or $G$-dwarf. Adding the additional priors for a full Bayesian analysis will alter the details of these contours but does not change the underlying discrepancy in the lens interpretation, which ultimately derives from the differences in the parallax contours.

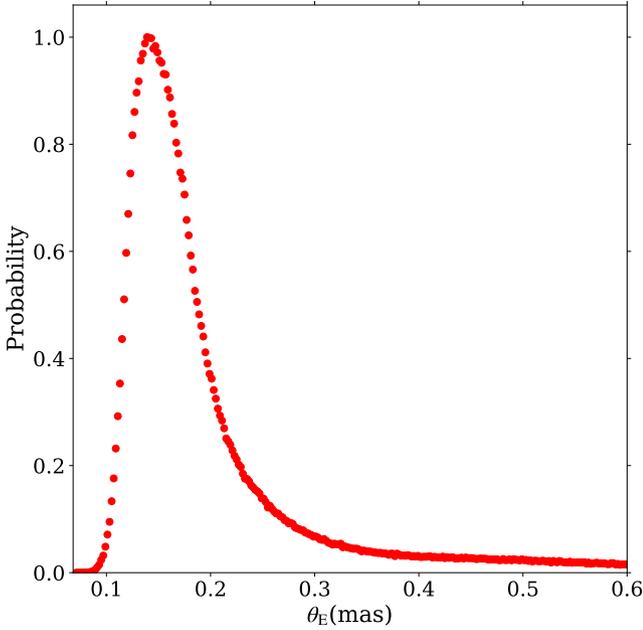

**Figure 8.** Probability distributions of the angular Einstein radius $\theta_E$, which is estimated by $\theta_E = \theta_*/\rho$. We obtain $\theta_*$ by CMD analysis (see Section 5.1), and $\rho$ is derived from the minimum $\chi^2$ for the lower envelope of the ($\chi^2$ vs. $\rho$) diagram from MCMC chain of full parallax models. $\theta_E = 0.14$ mas is the most-likely value and $\theta_E > 0.092$ mas at $3\sigma$ level, but the upper limit on $\theta_E$ is not constrained at the 2.7-$\sigma$ level.

### 5.3 Bayesian Analysis

We perform a Bayesian analysis using a Galactic model based on the mass function, stellar number density profile, and the source and lens velocity distributions. For the mass function of the lens, we choose the log-normal initial mass function of Chabrier (2003) and impose cut offs of 1.3 $M_\odot$ (Zhu et al. 2017b) and 1.1 $M_\odot$ (Bensby et al. 2017) for the disk lenses and bulge lenses, respectively. For the bulge and disk stellar number density profile, we choose the model used by Zhu et al. (2017b) and Bennett et al. (2014), respectively. For the source velocity distribution, we adopt the source proper motion measured by *Gaia* (Gaia Collaboration et al. 2016, 2018)

$$\vec{\mu}_S(N, E) = (-6.17 \pm 0.66, 0.54 \pm 0.74) \text{ mas yr}^{-1}. \quad (14)$$

For the velocity distribution of the lens in the Galactic bulge, we examine a *Gaia* CMD using the stars within 5 arcmin and derive the proper motion (in the Sun frame) for stars with $G < 18.5$; $B_P - R_P > 1.5$. We remove seven outliers and obtain

$$\langle \vec{\mu}_{\text{bulge}}(\ell, b) \rangle = (-5.9, -0.6) \pm (0.4, 0.3) \text{ mas yr}^{-1}, \quad (15)$$

$$\sigma(\vec{\mu}_{\text{bulge}}) = (2.7, 2.8) \pm (0.3, 0.3) \text{ mas yr}^{-1}. \quad (16)$$

Assuming the source distance is 7.55 kpc (inferred from the de-reddened brightness of the red clump $I_{cl} = 14.27$), the bulge stars toward this direction have mean velocity $\vec{v}(\ell, b) \sim (40, -10)$ km s$^{-1}$ and $\sigma_{\vec{v}} \sim 100$ km s$^{-1}$ velocity dispersion along each direction. For the disk lens velocity distribution, we assume the disk stars follow a rotation curve of 240 km s$^{-1}$ (Reid et al. 2014) and adopt the velocity dispersion of Han et al. (2020).

We create a sample of $2 \times 10^9$ simulated events and weight the six full parallax models shown in Table 4. For each simulated event $i$ of model $k$, the weight is given by

$$W_{\text{Gal},i,k} = \Gamma_{i,k} \mathcal{L}_{i,k}(t_E) \mathcal{L}_{i,k}(\vec{\pi}_E) \mathcal{L}_{i,k}(\theta_E), \quad (17)$$

where $\Gamma_{i,k} \propto \theta_{E,i,k} \times \mu_{\text{rel},i,k} \times D_L^2$ is the microlensing event rate, $\mathcal{L}_{i,k}(\vec{\pi}_E)$ and $\mathcal{L}_{i,k}(\theta_E)$ are the likelihood distribution for $\pi_E$ and $\theta_E$ shown in Figures 6 and 8, respectively, and $\mathcal{L}_{i,k}(t_E)$ are the likelihood of its inferred parameters $t_{E,i,k}$ given the error distributions of these quantities derived from the MCMC for that model

$$\mathcal{L}_{i,k}(t_E) = \frac{\exp[-(t_{E,i,k} - t_{E,k})^2/2\sigma_{t_E,k}^2]}{\sqrt{2\pi}\sigma_{t_E,k}}. \quad (18)$$

For each data set, we weight each solution by its probability for the Galactic model and $\exp(-\Delta\chi^2/2)$, where $\Delta\chi^2$ is the $\chi^2$ difference between the solution and the best-fit solution. In addition, the blended light is consistent with zero in $1\sigma$ (see Table 4), which can a provide useful constraint on the lens flux. We adopt 10% of the source flux as the upper limit of the lens flux, $I_{L,\text{limit}} = 19.9$, which is roughly the $3\sigma$ upper limit of the blended light. We then adopt the mass-luminosity relation of Wang et al. (2018),

$$M_I = 4.4 - 8.5 \log\left(\frac{M_L}{M_\odot}\right), \quad (19)$$

where $M_I$ is the absolute magnitude in the $I$ band, and reject trial events for which the lens properties obey

$$M_I + 5\log\frac{D_L}{10\text{pc}} + A_{I,D_L} < I_{L,\text{limit}}, \quad (20)$$

where $A_{I,D_L}$ is the extinction at $D_L$, which is derived by an extinction curve with a scale height of 120 pc and $A_{I,7.55\text{ kpc}} = 1.29$ from Nataf et al. (2013).

The distributions and relative weights for each solution and the combined results are shown in Table 5. For each solution, the resulting distributions of the lens host-mass $M_{\text{host}}$ and the lens distance $D_L$ are shown in Figures 10. The physical properties for the lens are different for different subsets of the *Spitzer* data. For "all" *Spitzer* data, the Bayesian analysis indicates that the lens system is composed of an $M_{\text{planet}} = 0.38^{+0.32}_{-0.16} M_J$ sub-Jupiter orbiting an $M_{\text{host}} = 0.13^{+0.12}_{-0.05} M_\odot$ M-dwarf or brown dwarf, the "early_2018 + 2019" subset suggests an $M_{\text{planet}} = 0.26^{+0.22}_{-0.11} M_J$ Saturn around an $M_{\text{host}} = 0.093^{+0.082}_{-0.038} M_\odot$ VLM dwarf, and the "2018" subset indicates an $M_{\text{planet}} = 1.65^{+0.71}_{-0.70} M_J$ Jupiter orbiting an $M_{\text{host}} = 0.60^{+0.26}_{-0.26} M_\odot$ more massive dwarf. The "early_2018 + 2019" subset prefers a disk planetary system, the "2018" subset prefers a bulge planetary system, and the "all" subset has an equal probability of a bulge or a disk system.

## 6 IMPLICATIONS

The difference between the parallax contours with and without the 2019 *Spitzer* baseline observations presents two problems. First, it complicates our interpretation of OGLE-2018-BLG-0799, because the different parallaxes result in radically different physical properties for the lens. Second, it is unclear whether or not this planet can be included in the statistical *Spitzer* sample for measuring the frequency of planets.

The goal of the *Spitzer* microlensing program is to create a statistical sample of events (including planets) with well-measured distances in order to probe variations in the frequency of planets along the line

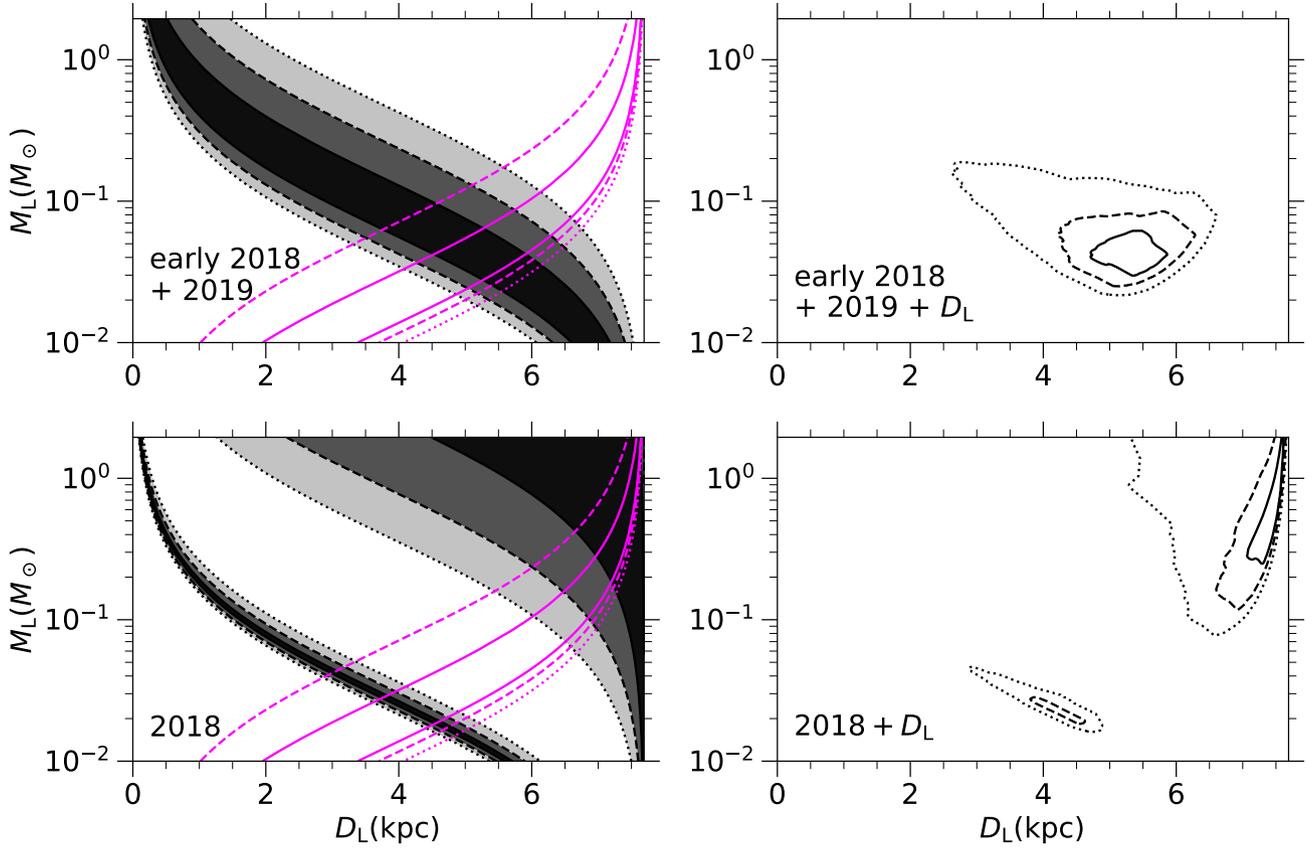

**Figure 9.** Constraints on the lens mass and system distance for OGLE-2018-BLG-0799. Left: 1-, 2-, and 3-$\sigma$ (solid, dashed, dotted) constraints from the angular Einstein radius $\theta_E$ (magenta) and the microlens parallax $|\pi_E|$ (black/shaded). Right: the joint constraint weighted by a factor $D_L^2$. Top: parallax constraint derived from the first half of the 2018 *Spitzer* data + the 2019 baseline observations ("early_2018 + 2019"). Bottom: parallax constraint derived from all 2018 *Spitzer* data but excluding the 2019 observations.

**Table 5.** Physical parameters from Bayesian analysis

| | | Physical Properties | | | | | | Relative Weights | |
|---|---|---|---|---|---|---|---|---|---|
| data set | Solutions | $M_{host}[M_\odot]$ | $M_{planet}[M_J]$ | $D_L$[kpc] | $r_\perp$[AU] | $\mu_{hel,rel}$[mas yr$^{-1}$] | $P_{bulge}$ | Gal.Mod. | $\chi^2$ |
| "all" | $u_{0,\oplus} > 0$ | $0.14^{+0.11}_{-0.05}$ | $0.38^{+0.31}_{-0.14}$ | $6.58^{+0.50}_{-0.48}$ | $1.19^{+0.33}_{-0.21}$ | $2.13^{+0.62}_{-0.38}$ | 0.979 | 0.728 | 0.301 |
| | $u_{0,\oplus} < 0$ | $0.13^{+0.12}_{-0.06}$ | $0.37^{+0.35}_{-0.17}$ | $5.29^{+1.30}_{-1.82}$ | $1.34^{+0.65}_{-0.35}$ | $2.97^{+3.52}_{-1.06}$ | 0.395 | 1.000 | 1.000 |
| | combined | $0.13^{+0.12}_{-0.05}$ | $0.38^{+0.32}_{-0.16}$ | $6.28^{+0.64}_{-2.02}$ | $1.24^{+0.52}_{-0.26}$ | $2.33^{+2.22}_{-0.52}$ | 0.500 | ... | ... |
| "2018" | $u_{0,\oplus} > 0$ | $0.62^{+0.25}_{-0.25}$ | $1.68^{+0.68}_{-0.68}$ | $7.14^{+0.50}_{-0.50}$ | $1.41^{+0.49}_{-0.29}$ | $2.27^{+0.82}_{-0.46}$ | 0.979 | 0.733 | 1.000 |
| | $u_{0,\oplus} < 0$ | $0.59^{+0.27}_{-0.26}$ | $1.62^{+0.72}_{-0.71}$ | $7.06^{+0.54}_{-0.56}$ | $1.45^{+0.59}_{-0.32}$ | $2.35^{+1.04}_{-0.52}$ | 0.896 | 1.000 | 0.905 |
| | combined | $0.60^{+0.26}_{-0.26}$ | $1.65^{+0.71}_{-0.70}$ | $7.10^{+0.52}_{-0.54}$ | $1.43^{+0.55}_{-0.31}$ | $2.31^{+0.94}_{-0.48}$ | 0.933 | ... | ... |
| "early_2018 + 2019" | $u_{0,\oplus} > 0$ | $0.074^{+0.030}_{-0.020}$ | $0.21^{+0.08}_{-0.06}$ | $6.19^{+0.47}_{-0.43}$ | $1.02^{+0.20}_{-0.15}$ | $1.95^{+0.42}_{-0.28}$ | 0.980 | 0.367 | 0.183 |
| | $u_{0,\oplus} < 0$ | $0.096^{+0.083}_{-0.040}$ | $0.27^{+0.23}_{-0.12}$ | $3.93^{+3.16}_{-1.84}$ | $1.31^{+0.47}_{-0.36}$ | $3.93^{+3.16}_{-1.84}$ | 0.147 | 1.000 | 1.000 |
| | combined | $0.093^{+0.082}_{-0.038}$ | $0.26^{+0.22}_{-0.11}$ | $4.05^{+1.78}_{-1.16}$ | $1.28^{+0.48}_{-0.34}$ | $3.71^{+3.24}_{-1.70}$ | 0.199 | ... | ... |

[1] $P_{bulge}$ is the probability of a lens in the Galactic bulge. The combined result of each *Spitzer* data set is obtained by a combination of $u_{0,\oplus} > 0$ and $u_{0,\oplus} < 0$ solutions weighted by the probability for the Galactic model and $\exp(-\Delta\chi^2/2)$.

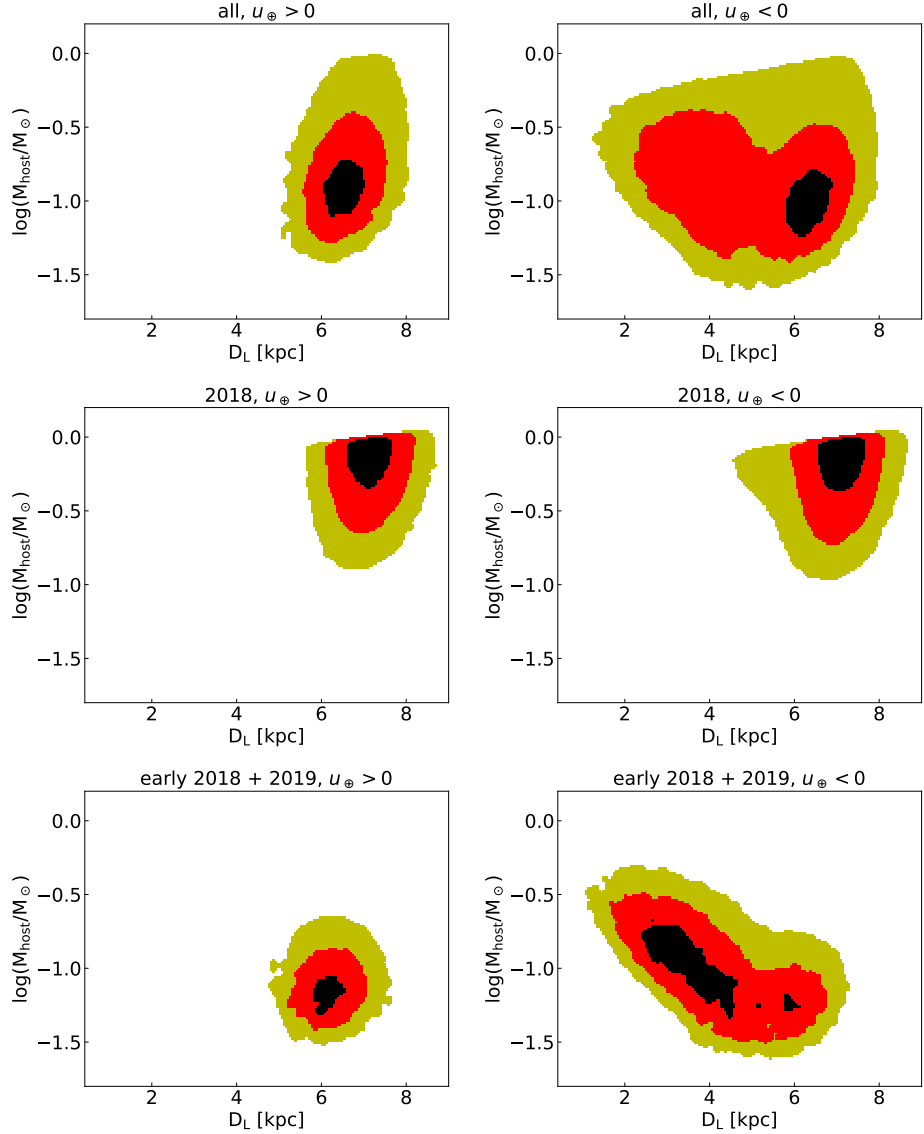

**Figure 10.** Bayesian posteriors distributions of the lens host-mass $M_{\rm host}$ and the lens distance $D_{\rm L}$. In each panel, black, red, and yellow colors show likelihood ratios $[-2\Delta \ln \mathcal{L}/\mathcal{L}_{\rm max}] < (1, 4, 9)$, respectively.

of sight. Previously, Zhu et al. (2017b) proposed that events should have

$$\sigma(D_{8.3}) < 1.4 \text{ kpc}; \qquad D_{8.3} \equiv \frac{\text{kpc}}{1/8.3 + \pi_{\rm rel}/\text{mas}} \qquad (21)$$

to be included in the sample. For a planetary event, $\sigma(D_{8.3})$ should be evaluated based on data from which the planet has been removed and only including *Spitzer* data scheduled without knowledge of the planet (so that the event can be evaluated under the same conditions as events without planets). We follow the procedures described in Ryu et al. (2018) to fit with a point-lens model using the analogous data and conduct a Bayesian analysis without the constraint of the finite-source effects. We find $D_{8.3} = 3.72^{+1.42}_{-1.00}$ kpc for the "all" *Spitzer* data, $D_{8.3} = 7.40^{+0.45}_{-0.81}$ kpc for the "2018" subset and $D_{8.3} = 3.09^{+0.91}_{-0.79}$ kpc for the "early_2018 + 2019" subset. So $D_{8.3}$ is constrained well enough at 1-$\sigma$ to meet the Zhu et al. (2017b) crite-rion in two of three cases, and especially in the "2018" case by which the criterion should be evaluated. However, the parallax as measured from the 2018 *Spitzer* data alone is different from the parallax based on an analysis including the 2019 *Spitzer* baseline data. Furthermore, in the case with "all" data, the constraints on $D_{8.3}$ are worse and fail the criterion. This suggests that we may need to re-evaluate how we interpret parallaxes measured from *Spitzer* light curves and also how the statistical sample of *Spitzer* events is defined.

The change in the parallax contours with the addition of 2019 *Spitzer* baseline observations indicate that systematics in the photometry are affecting the parallax constraint. Some level of systematics (or rather correlated noise) has always been present in the *Spitzer* photometry of microlensing events (e.g., Poleski et al. 2016). As noted in Zhu et al. (2017b), there are several examples of cases for which the annual parallax effect confirms the satellite parallax effect (Udalski et al. 2015b; Han et al. 2017). Hence, Zhu et al. (2017b) concludes that these systematics do not have a significant effect on the resulting parallax measurements. By contrast, Koshimoto & Bennett (2020) compared the parallaxes measured for the Zhu et al. (2017b) sample to a predicted distribution of parallaxes from a galactic model. Based on the differences between the observed parallaxes and their prediction, they concluded that systematics caused *Spitzer* parallaxes to be overestimated. However, they did not investigate the actual *Spitzer* photometry.

OGLE-2018-BLG-0799 shows that, for at least some events, systematics in the photometry does play a significant role in the measured parallaxes. Thus, this issue requires a more systematic investigation of the photometry (and the resulting constraints on the parallax) in order to understand how often systematics in the photometry affect the measured parallax, the conditions under which those problems appear, and how the parallax measurement is affected.

The arc-like form of the parallax contours in OGLE-2018-BLG-0799 suggests the work of Gould (2019) can offer a deeper understanding of how to robustly assess the satellite parallaxes in the presence of photometric systematics. The development of the *Spitzer*-"ONLY" method for investigating the satellite parallax has shown that the uncertainty contours for the parallax measured in the $\pi_{E,E}$-$\pi_{E,N}$ plane are frequently arc-shaped rather than simple ellipses (Shin et al. 2018; Jung et al. 2019; Gould et al. 2020; Zang et al. 2020a,b; Hirao et al. 2020). Gould (2019) then showed the theoretical origin of these arcs. Given a color-constraint and a measurement of the baseline flux, each *Spitzer* observation yields a circular constraint on the parallax. Then, when combined, a group of late-time observations yields a series of osculating circles whose intersection defines the measurement of the parallax.

A partial ring (as would be created by a series of osculating circles) is exactly the form of the constraint that we see for OGLE-2018-BLG-0799. This suggests that the 2018 data alone give a good measurement of the resulting arc, but the systematics in this event lead to the wrong localization along this arc. Future investigations of the influence of systematics in *Spitzer* photometry on the measured parallaxes should focus on further understanding at these arc-like constraints and their relationship to the osculating circles of Gould (2019). In addition, the criterion for assessing membership in the statistical *Spitzer* sample may need to be revised to account for these arcs and the two-dimensional nature of the parallax constraints.

## 7 CONCLUSION

In this paper, we have reported the discovery and analysis of the *Spitzer* microlens planet OGLE-2018-BLG-0799Lb. The mass ratio between the lens star and its companion is $q = (2.65 \pm 0.16) \times 10^{-3}$. The combined constraints from $\theta_E$ and $\pi_E$ suggest that the host star is most likely to be a very low-mass dwarf. In our preferred solution using the subset of the *Spitzer* data from "early_2018 + 2019", a full Bayesian analysis indicates that the planetary system is composed of a $M_{\rm planet} = 0.26^{+0.22}_{-0.11}$ $M_J$ planet orbiting a $M_{\rm host} = 0.093^{+0.082}_{-0.038}$ $M_\odot$ dwarf, with a host-planet projected planet separation $r_\perp = 1.28^{+0.48}_{-0.34}$ AU, which indicates that the planet is a Saturn-mass planet beyond the snow line of a very low-mass dwarf (assuming a snow line radius $r_{\rm SL} = 2.7(M/M_\odot)$ AU, Kennedy & Kenyon 2008). However, because of systematics in the *Spitzer* photometry, there is ambiguity in the parallax measurement. Using all of the *Spitzer* data yields a parallax that implies $M_{\rm host} = 0.13^{+0.12}_{-0.05}$ $M_\odot$ at $D_L = 6.28^{+0.64}_{-2.02}$ kpc. Although we consider a very low-mass object in the disk to be the most likely explanation for the host star, it is also possible for it to be a more massive star in the Galactic bulge. Indeed, in the absence of the 2019 data, we would have concluded $M_{\rm host} = 0.60^{+0.26}_{-0.26}$ at $D_L = 7.10^{+0.52}_{-0.54}$ kpc.

An adaptive optics measurement of (or constraint on) the lens flux would substantially improve the constraints on the lens and distinguish between the different parallax solutions. A strong upper limit on the flux could immediately rule out the 2018-only solution, and a detection would be constraining although some ambiguities may persist due to potential confusion with other stars. Furthermore, if one waited until the lens and source could be separately resolved (e.g., Bhattacharya et al. 2020), if the lens were detected, this would yield a measurement of the lens-source relative proper motion vector, $\vec{\mu}_{\rm rel}$. Its magnitude, $|\vec{\mu}_{\rm rel}|$, would give a measurement of $\theta_E$, which is only constrained by the microlensing light curve. In addition, a measurement of $\hat{\vec{\mu}}_{\rm rel} = \hat{\vec{\pi}}_E$ would further constrain the parallax contours (e.g., Zang et al. 2020b), both improving the measurement of $\vec{\pi}_E$ and independently testing the impact of systematics in the *Spitzer* photometry. The lens-source relative proper motion in this event is slow ($\mu_{\rm rel} \sim 3$ mas yr$^{-1}$), but such a measurement could be made in $\sim 20$ years with a 8-10m class telescope if the lens is luminous or at first light of AO imagers on 30m telescopes (or possibly with JWST) if the lens is a faint brown dwarf.

### ACKNOWLEDGEMENTS

W.Z., X.Z., H.Y. and S.M. acknowledge support by the National Science Foundation of China (Grant No. 12133005). Work by JCY was supported by JPL grant 1571564. The OGLE has received funding from the National Science Centre, Poland, grant MAESTRO 2014/14/A/ST9/00121 to AU. This research has made use of the KMTNet system operated by the Korea Astronomy and Space Science Institute (KASI) and the data were obtained at three host sites of CTIO in Chile, SAAO in South Africa, and SSO in Australia. The MOA project is supported by JSPS KAKENHI Grant Number JSPS24253004, JSPS26247023, JSPS23340064, JSPS15H00781, JP16H06287, and JP17H02871. This research uses data obtained through the Telescope Access Program (TAP), which has been funded by the National Astronomical Observatories of China, the Chinese Academy of Sciences, and the Special Fund for Astronomy from the Ministry of Finance. This work is based (in part) on observations made with the *Spitzer* Space Telescope, which is operated by the Jet Propulsion Laboratory, California Institute of Technology under a contract with NASA. Support for this work was provided by NASA through an award issued by JPL/Caltech. Work by AG was supported by AST-1516842 and by JPL grant 1500811. AG received support from the European Research Council under the European Unions Seventh Framework Programme (FP 7) ERC Grant Agreement n. [321035]. Wei Zhu was supported by the Beatrice and Vincent Tremaine Fellowship at CITA. Work by CH was supported by the grants of National Research Foundation of Korea (2017R1A4A1015178 and 2019R1A2C2085965). YT acknowledges the support of DFG priority program SPP 1992 Exploring the Diversity of Extrasolar Planets (WA 1047/11-1). This research has made use of the NASA Exoplanet Archive, which is operated by the California Institute of Technology, under contract with the National Aeronautics and Space Administration under the Exoplanet Exploration Program.

## DATA AVAILABILITY

Data used in the light curve analysis will be provided along with publication.

This paper has been typeset from a T<sub>E</sub>X/L<sup>A</sup>T<sub>E</sub>X file prepared by the author.